\documentclass[pra,twocolumn,superscriptaddress,showpacs]{revtex4}
\usepackage{graphicx}       
\usepackage{bm}         
\usepackage{amsmath}        
\usepackage{latexsym}
\usepackage{epsfig}

\newcommand{\beq}{\begin{equation}}
\newcommand{\eeq}{\end{equation}}
\newcommand{\beqa}{\begin{eqnarray}}
\newcommand{\eeqa}{\end{eqnarray}}

\newcommand{\non}{\nonumber}

\begin{document}

\title{Cavity state preparation using adiabatic transfer}
\author{Jonas Larson}
\affiliation{Physics Department, Royal Institute of Technology (KTH), Albanova, Roslagstullsbacken 21, SE-10691 Stockholm, Sweden}
\author{Erika Andersson}
\affiliation{Department of Physics, John Anderson Building, University of Strathclyde, Glasgow G4 0NG, UK}
\begin{abstract}
We show how to prepare a variety of cavity field states for multiple cavities. The state preparation technique used is related to the method of stimulated adiabatic Raman passage or STIRAP. The cavity modes are coupled by atoms, making it possible to transfer an arbitrary cavity field state from one cavity to another, and also to prepare non-trivial cavity field states.
In particular, we show how to prepare entangled states of two or more cavities, such as an $EPR$ state and a $W$ state, as well as various entangled superpositions of coherent states in different cavities, including Schr\"odinger  cat states. The theoretical considerations are supported by numerical simulations. 
\end{abstract}
\pacs{42.50.Dv, 32.80.-t, 03.67.Mn, 03.67.-a
 }
\maketitle

\section{Introduction}\label{s1}
A recent paper \cite{fredrik} presented an efficient method to adiabatically transfer field states between two different cavities. The scheme is closely related to {\it stimulated Raman adiabatic passage}, or shortly {\it STIRAP} \cite{stirap,review}. STIRAP was first used to coherently control dynamical processes in atoms and molecules. Two external laser pulses drive population between an initial and a final state in an atom or molecule, through an intermediate level. One pulse couples the initial state to the intermediate state and the other pulse couples the intermediate and final state. The pulses are applied in a counterintuitive way, in the sense that  the pulse that couples the final and intermediate states is turned on first. The pulses do have to overlap though,  and in order for the process to work successfully it has to be adiabatic, as the name suggests. Population will then follow the instantaneous eigenstates adiabatically. One of the eigenstates is of particular interest, namely the {\it dark state}. This state has eigenvalue zero, and the intermediate state is never populated during the evolution. 

In the method suggested in \cite{fredrik}, a two-level atom interacts with two cavities. In this scheme, the  couplings between the atom and the two cavities correspond to the the two laser pulses in traditional STIRAP. As the atom traverses the cavities it will see the varying shape of the mode it interacts with, and consequently, the coupling becomes time-dependent. By letting the cavities partly overlap spatially, it is possible to realize a situation very similar to STIRAP. In fact, if the state, adiabatically transmitted between the cavities, is a one photon state $|1\rangle$, the corresponding Hamiltonian (in the dipole and rotating wave approximations) looks exactly the same as the standard STIRAP one. The ingenious feature of the method is that it works for any field state, not just the one photon state. The Hilbert space will, of course, increase when larger photon number states are involved, and therefore the adiabaticity constraints become more stringent \cite{fredrik2}. There is still a dark state with zero population in the upper atomic level, even for general field states. 

Other schemes, where the atom experiences a varying mode shape as it traverses the cavity, have also been suggested for adiabatic state preparation of the field modes  \cite{ent1, agarwal,ent2, ent3, jonas}. However, these schemes differ from the present model. For example, in papers \cite{ent1,agarwal} a lambda type atom is used, in \cite{ent1,ent2,ent3} a strong external classical laser field is utilized and in \cite{jonas} only one cavity and one two-level atom is considered.

In this paper we will extend the model in \cite{fredrik} to more complex systems involving more than just one two-level atom and two cavities. As we have mentioned, in the one photon case the model in \cite{fredrik} is analogous with the traditional STIRAP. Likewise, the extensions made in this paper are related to similar generalizations of the traditional STIRAP, if we consider the one photon case. General situations for multi-level STIRAP has been analyzed in several papers; just to mention a few, see \cite{multistirap, multistirap1, multistirap2, multistirap3, multistirap4}. By including more atoms and cavities, we will show that various interesting field states can be prepared. Due to the fact that the dimension of the accessible Hilbert space easily blows up when the photon number is increased in these extended models, we will choose the transferred field state  to contain just one photon in our numerical simulations. However, in the adiabatic limit, the system is solvable also for higher photon numbers. Using more photons  only means that the adiabaticity constraints are stricter, as mentioned above. As compared with the method in \cite{fredrik}, we will note that also these more complicated systems have an adiabatic dark state, which will be used for the evolution. It will be shown that it is possible to entangle spatially separated cavities, and prepare, for example, $EPR$ or $W$ field states, but also more complex entangled states. By making atomic measurements, it is feasible to create Schr\"odinger cat states. The setups given in this paper are only a couple of examples, and others are of course possible; we just illustrate the basic idea. We consider preparation of the various field states, but the methods could equally well be applied for creating different atomic states if desired.

The outline of the paper is as follows: In section \ref{s2} we review the basic idea and properties of the method presented  in \cite{fredrik}. We introduce the adiabatic eigenstates and explain the dynamics behind the transfer of arbitrary field states between two cavities. In section \ref{s3} we consider two different setups, which we call the ``H" configuration, consisting of three cavities and the "star" configuration, which could contain any number $M$ of cavities. In the H configuration we show how a state is transfered between two spatially separated cavities by virtual pass through a third cavity and it is also explained how $EPR$ states could be prepared. The other model, the star configuration, could also be used for achieving $EPR$ states as well as $W$ states and generalizations of these states. In section \ref{s4}, we make use of a third atomic level and projective atomic measurements for preparing various types of Schr\"odinger cat states. Finally we conclude with a summary and discussion in section \ref{s5}. 

\section{Adiabatic transfer between cavity modes}\label{s2}

We will first briefly review how to adiabatically transfer a quantum state from one cavity mode to another, following \cite{fredrik}. We consider a situation where there are two cavity modes interacting with a single two-level atom. The Hamiltonian for this system is a generalisation of the widely used Jaynes-Cummings model \cite{jaynescummings},
\begin{eqnarray}
H&=&\frac{1}{2} \omega(\sigma_{z}+1)+\Omega_1\hat a_1^\dagger \hat a_1+\Omega_2\hat a_2^\dagger \hat a_2 \non\\
&+&(g_{1}\hat a_1 + g_{2}\hat a_2)\sigma^+_a+(g_{1}\hat a^\dagger_1 + g_{2}\hat a^\dagger_2)\sigma^-_a.
\end{eqnarray}
Here $\hat a^\dagger_1$ and $\hat a^\dagger_2$ are the boson creation operators for cavity modes 1 and 2, respectively, $\sigma_z, \sigma^+$ and $\sigma^-$ are the Pauli $z$ and the raising and lowering operators for the atom, and $g_1(t)$ and $g_2(t)$ describe the time-dependent coupling between the light and the two-level atom. The basis states for the system are of the form
\begin{equation}
|n_1,n_2,s\rangle \equiv |n_1\rangle |n_2\rangle |s\rangle,
\end{equation}
where $n_1$ and $n_2$ refer to the number of excitations in mode 1 and 2, and $s=\pm$ refers to the state of the two-level atom, with $\sigma_z |s\rangle = s|s\rangle$. In the following we will assume that the cavity modes are degenerate, $\Omega_1=\Omega_2=\Omega$, so that perfect transfer of excitations between the modes is possible. If we start with a single excitation in mode 1 and the atom in its ground state, then the accessible Hilbert space is spanned by the three states
\begin{equation}
|1,0,-\rangle, |0,0,+\rangle, |0,1,-\rangle.
\end{equation}
The Hamiltonian commutes with the operator
\begin{equation}
N=\frac{1}{2}(\sigma_z+1)+\hat a^\dagger_1\hat a_1+\hat a^\dagger_2\hat a_2,
\end{equation}
so that we can work in an interaction picture, with the Hamiltonian
\begin{eqnarray}
H'&=&H-\Omega N \\
&=& \Delta (\sigma_z+1)+[(g_1(t)\hat a_1+g_2(t)\hat a_2)\sigma^+ +h.c.],\non
\end{eqnarray}
where $\Delta=(\omega-\Omega)/2$. The atom does not need to be on resonance with the cavity modes, i.e. $\Delta$ can be nonzero.

As in the case of adiabatic transfer between atomic states \cite{stirap,review,timostirap}, there is an eigenstate of this Hamiltonian with eigenvalue zero, given by
\begin{equation}
|\Psi_{ad}\rangle=K_{12}\left[g_2(t)|1,0,-\rangle - g_1(t)|0,1,-\rangle\right],
\label{adstate}
\end{equation}
where the normalisation constant is given by $K_{12}^{-2}=g_1^2(t)+g_2^2(t)$. Consider the case when
\begin{eqnarray}
\lim_{t\rightarrow -\infty}\frac{g_1(t)}{g_2(t)}&=&0\nonumber\\
\lim_{t\rightarrow \infty}\frac{g_2(t)}{g_1(t)}&=&0.
\label{stirapcond}
\end{eqnarray}
If the couplings $g_1(t)$ and $g_2(t)$ change slowly enough, the system will start in the state $|1,0,-\rangle$, and end up in the state $|0,1,-\rangle$, following the adiabatic eigenstate given in equation (\ref{adstate}). This method is called stimulated Raman adiabatic passage or STIRAP  \cite{stirap,review}. The exact shapes of the pulses $g_1(t)$  and $g_2(t)$ do not matter, as long as they vary slowly enough  and conditions (\ref{stirapcond}) hold. The pulse sequence is {\it counterintuitive} in the sense that the two initially empty levels are coupled first, and only then is the initially populated level coupled to the ``middle" level. The two pulses $g_1(t)$ and $g_2(t)$ must, however, overlap.

By choosing $\lim_{t\rightarrow \infty}g_2(t)/g_1(t) = 1$ instead of 0, we can also adiabatically reach the state
\begin{equation}
\frac{1}{\sqrt{2}}(|1,0,-\rangle -|0,1,-\rangle),
\end{equation} 
or, by choosing another suitable ratio between $g_1(t\rightarrow\infty)$ and $g_2(t\rightarrow\infty)$, we can reach any superposition of $|1,0,-\rangle$ and $|0,1,-\rangle$. This process is referred to as {\it fractional STIRAP} \cite{review}. 

\subsection{Transfer of an arbitrary cavity field state}\label{ss21}

Also more than one field excitation can be transferred between the cavity modes \cite{fredrik}. For example, a Fock state $|n\rangle$ in mode 1 can be transferred to mode 2. We can write the adiabatic state (\ref{adstate}) as
\begin{equation}\label{adop}
|\Psi_{ad}\rangle = \hat A^\dagger|0,0,-\rangle,
\end{equation}
where the boson operator $\hat A^\dagger$ is defined as
\begin{equation}
\hat A^\dagger = K_{12}(g_2\hat a^\dagger_1-g_1\hat a^\dagger_2).
\end{equation}
The Hamiltonian, on the other hand, can be written as
\begin{equation}
H' = \Delta(\sigma_z+1)+K_{12}^{-1}(\hat B \sigma^++\hat B^\dagger \sigma^-),
\end{equation}
where the boson operator $\hat B^\dagger$ is given by
\begin{equation}
\hat B^\dagger = K_{12}(g_1\hat a^\dagger_1+g_2 \hat a^\dagger_2).
\end{equation}
We find that $[\hat B,\hat A^\dagger] =0$, so that the state
\begin{equation}
|\Psi^n_{ad}\rangle = \frac{1}{(n!)^{1/2}}(\hat A^\dagger)^n|0,0,-\rangle
\end{equation}
is an adiabatic state, since $H'|\Psi^n_{ad}\rangle = 0$. Choosing the couplings so that conditions (\ref{stirapcond}) hold, we immediately find that the state $|n,0,-\rangle$ adiabatically changes into $|0,n,-\rangle$.

More generally, we can consider the adiabatic state
\begin{equation}
f(\hat A^\dagger)|0,0,-\rangle = C_n \frac{(\hat A^\dagger)^n}{(n!)^{1/2}}|0,0,-\rangle.
\end{equation}
If the couplings again satisfy conditions (\ref{stirapcond}), and if we choose the pulses so that $g_1/g_2 <0$, then the state $f(\hat a^\dagger_1)|0,0,-\rangle$ will adiabatically change into $f(\hat a^\dagger_2)|0,0,-\rangle$. For example, a coherent state $|\alpha\rangle$ can be transferred from cavity mode 1 to cavity mode 2 by choosing 
\begin{equation}
|\Psi_{ad}\rangle = \exp\left(-\frac{|\alpha |^2}{2}\right)\exp(\alpha \hat A^\dagger)|0,0,-\rangle.
\end{equation}

\section{Adiabatic transfer with multiple cavities}\label{s3}

\subsection{Three cavities and two atoms in an ``H" configuration}\label{ss31}

\begin{figure}[!htb]
\center{\includegraphics[width=8cm,height=!]{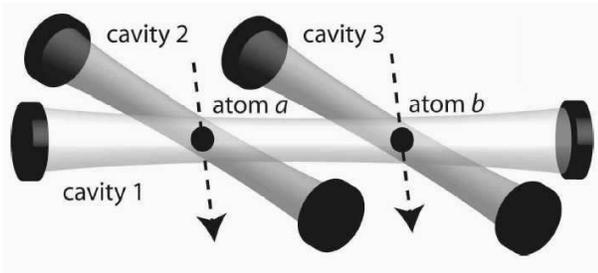}}
\caption{A possible setup of the three cavities ($1$, $2$ and $3$) and the two atomic ($a$ and $b$) trajectories for the ``H configuration". } 
\label{fig1}
\end{figure}

We will now move on to consider cavity state transfer in a situation where we have three cavities and two atoms.
Suppose cavities 1, 2 and 3 are placed so that cavity 1 is overlapping with both cavities 2 and 3. Atom $a$ is placed in the crossing between cavities 1 and 2, and atom $b$ in the crossing between cavities 2 and 3, as shown in figure \ref{fig1}. The Hamiltonian for this system is given by
\begin{equation}
\begin{array}{ccl}
H&=&\frac{1}{2} \omega_a(\sigma_{az}+1)+\frac{1}{2}\omega_b(\sigma_{bz}+1)
+\Omega_1\hat a_1^\dagger \hat a_1\\ \\ &+&\Omega_2\hat a_2^\dagger \hat a_2+\Omega_3\hat a_3^\dagger \hat a_3+\left[(g_{1a}\hat a_1 + g_{2a}\hat a_2)\sigma^+_a\right.\\ \\ & + & \left.(g_{1b}\hat a_1 + g_{3b}\hat a_3)\sigma^+_b+h.c\right],
\end{array}
\end{equation}
where $\sigma_{a(b)z}, \sigma_{a(b)}^+$  and $\sigma_{a(b)}^-$ refer to atom $a(b)$, and $\hat a_i^\dagger$ and $\hat a_i$ are the creation and annihilation operators for cavity $i$. We have denoted the coupling strengths between cavity $i$ and atom $a$ as $g_{ia}$, and correspondingly for atom $b$. The number of excitations in the systems is conserved, and we find that the Hamiltonian commutes with the operator
\beq
N=\frac{1}{2}(\sigma_{az}+1)+\frac{1}{2}(\sigma_{az}+1)+\hat a_1^\dagger \hat a_1+\hat a_2^\dagger \hat a_2+\hat a_3^\dagger \hat a_3.
\eeq
In the following we will assume that $\Omega_1=\Omega_2=\Omega_3\equiv\Omega$. Otherwise perfect transfer of cavity field states would not be possible, since energy is conserved. In the interaction picture, we form the Hamiltonian
\begin{equation}
\begin{array}{ccl}
\widetilde{H}&=&H-\Omega N=\Delta_a(\sigma_{az}+1)+\Delta_b(\sigma_{bz}+1)\\ \\
&+&\left[(g_{1a}\hat a_1 + g_{2a}\hat a_2)\sigma^+_a+(g_{1b}\hat a_1 + g_{3b}\hat a_3)\sigma^+_b+h.c.\right],
\end{array}
\end{equation}
where $\Delta_a=(\omega_a-\Omega_a)/2$, and similarly for $b$. We now write the basis states as $|n_1, n_2, n_3, \pm_a,\pm_b\rangle$, where the three first entries refer to the number of photons in cavities 1, 2 and 3, and the two last entries to the states of the atoms. The subspaces with exactly one excitation in the system is spanned by the five basis states $|0,1,0,-,-\rangle, |0,0,0,+,-\rangle, |1,0,0,-,-\rangle, |0,0,0,-,+\rangle$ and $|0,0,1,-,-\rangle$. Using this ordering of the basis states, the Hamiltonian in matrix form for this subspace becomes
\beq
\widetilde{H}=\left(\begin{array}{c c c c c}
0              & g_{2a}     &0              & 0              & 0\\
g^*_{2a} & \Delta_a & g_{1a}    & 0              & 0\\
0 	       & g^*_{1a} & 0              & g_{1b}    & 0\\   
0              & 0              & g^*_{1b} & \Delta_b & g_{3b}\\
0              & 0              & 0              & g^*_{3b} & 0
\end{array}\right).
\eeq
This Hamiltonian has an  adiabatic eigenstate  with eigenvalue zero. Making the Ansatz $(C_2, 0, C_1, 0, C_3)^{\rm T}$ for this state, the condition on the coefficients $C_i$ becomes $g^*_{2a}C_2+g_{1a}C_1 = g^*_{1b}C_1+g_{3b}C_b = 0$, so that the adiabatic eigenstate is
\beq\label{hadiabatic}
|\Psi\rangle_{ad}=K( g_{1a} g_{3b},  0,  -g^*_{2a} g_{3b}, 0, g^*_{1b} g^*_{2a})^{\text T},
\eeq
where $K$ is a normalisation constant.
We see that there should be a possibility of transferring the state of cavity 2 directly to cavity 3 with very little population in cavity 1. For a thorough exposition of adiabatic transfer between atomic levels with multiple intermediate states, see \cite{multistirap}. The theory can be directly applied to cavity state transfer as well. To achieve transfer from cavity 2 to cavity 3, we should start with 
\begin{equation}
\label{startcond}
|g_{1a}g_{3b}|\gg |g_{1b}g_{2a}|, 
\end{equation}
and finish with 
\begin{equation}
\label{endcond}
|g_{1b}g_{2a}|\gg |g_{1a}g_{3b}|,
\end{equation}
keeping
\begin{equation}
\label{alltimescond}
|g_{1a}g_{3b}|^2+ |g_{1b}g_{2a}|^2\gg |g_{2a}g_{3b}|^2
\end{equation}
all the time. There are many possible pulse sequences satisfying these conditions. 
A few possible coupling sequences  will be discussed in the next subsection. In all cases we start with one field excitation in cavity 2. 

As for the case where two cavity modes are coupled by one atom \cite{fredrik}, the transfer of arbitrary cavity states from mode 2 to mode 3 will also be possible. If we form the ``adiabatic operator"
\beq
\small
\hat A^\dagger(t)=\!K(t)\!\left[{g_{1a}(t) {g_{3b}(t)}} \hat a^\dagger_2 -g^*_{2a}(t) g_{3b}(t)\hat a^\dagger_1+{g^*_{1b}(t) {g^*_{2a}(t)}}\hat a^\dagger_3\right],
\eeq 
where $K(t)$ is a normalisation constant,
then,  in the adiabatic limit, if we start in the state $f[\hat A^\dagger(0)]|0\rangle$, we will also stay in the state $f[\hat A^\dagger(t)]|0\rangle$ as the couplings are changed. For example, starting in $f(\hat a_2^\dagger )|0,0,-\rangle$, we can adiabatically transfer the cavity state to mode 3,  $f(\hat a_3^\dagger )|0,0,-\rangle$. As before, this means that we can transfer not only one field excitation, but also, for example, number states, where $f(A^\dagger)=A^{\dagger n}$, and coherent states, where $f(A^\dagger)=\exp\left(|\alpha |^2/2\right)\exp(\alpha \hat A^\dagger)$.

\subsection{Numerical simulations of the ``H" configuration}\label{ss32}
For all the numerical simulations in the paper we use Gaussian pulses for the couplings, of the form
\begin{equation}
g_{i\nu}(t)=G_{i\nu}\exp\left(-\frac{(t-t_{i\nu})^2}{\sigma_{i\nu}^2}\right).
\end{equation}
The index $i$ stands for the $i$'th cavity and $\nu$ for atom $\nu$; cavities will be labeled with numbers and atoms with letters. If there is only one atom present the atomic index will be omitted. $G$ is the coupling amplitude, and it will be chosen the same for all pulses in the different examples, except for a couple of examples in the next section. The indices will be omitted when the $G$:s are all the same. The parameter $t_{i\nu}$ gives the pulse center and the width is given by $\sigma_{i\nu}$. We are using scaled parameters with $\hbar =1$. Time $t$ and the pulse widths $\sigma$ are given in units of a suitable characteristic time $T$, and $G$ and $\Delta$ in units of $\hbar T^{-1}$.

\begin{figure}
\center{\includegraphics[width=8cm,height=!]{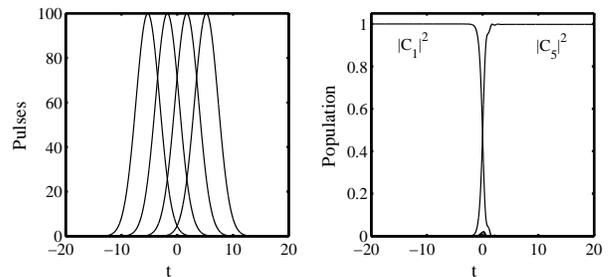}}
\caption {The figure to the left shows our first example of a pulse sequence for realizing complete population transfer from cavity 2 to cavity 3 with minimal population in the intermediate cavity 1 for the "H" configuration. The pulses are ordered in a completely counterintuitive way, from left to right $g_{3b}$, $g_{1b}$, $g_{1a}$ and $g_{2a}$. Time is given in units of a suitable characteristic time $T$. The widths of the pulses are all $\sigma=3$, also in units of $T$, and the maximum amplitudes are $G=100$ in units of $T^{-1}$. The other plot shows the populations $|C_i|^2$ ($i=1,\, 2,\, 3,\, 4$ or $5$) as a function of the scaled interaction time $t$. It is clear that population is transfered adiabatically from the second cavity (solid line marked $|C_1|^2$) to the third cavity (dotted line marked $|C_5|^2$), without remarkable population in cavity 1. The final population in the third cavity is 99.8 $\%$, and maximum population of cavity 1 during the process is 0.2 $\%$. 
} 
\label{fig2}
\end{figure}

We will consider two possible pulse sequences for adiabatic transfer in the "H" configuration. The first pulse sequence, which is shown in figure \ref{fig2}, is completely counter-intuitive, in the sense that we start by coupling cavity 3 and atom $b$, then  cavity 1 and atom $b$, followed by cavity 1 and atom $a$, and finally cavity 2 and atom $a$. This could for example be achieved if the cavities are crossing each other horizontally, partly overlapping, and we let the atom $b$ traverse first cavity 3 and then cavity 1, and similarly for atom $a$ and cavities 1 and 2. The parameters in the figure are $t_{3b}=-5.22$, $t_{1b}=-1.72$, $t_{1a}=1.78$ and $t_{2a}=5.28$, $\sigma=3$, $\Delta=0$ and $G=100$. The dynamics is, for $\Delta = 0$, determined by the dimensionless adiabaticity parameter $G\sigma$ \cite{fredrik}.
 
The pulses are seen in the left plot and the populations in the right one.
As shown in figure \ref{fig2}, numerical simulations confirm that an excitation in cavity 2 can be transferred adiabatically to cavity 3, while  the population in cavity 1 remains small in between. The final population in state $|0,0,1,-,-\rangle$ is 99.8~\% and maximum population in cavity 1 is 0.2~\% and is located around $t=0$. The coupling amplitudes are rather large in this example in order to have an adiabatic process and correspondingly a successful transfer. This is due to the fact that the population virtually passes through three levels, $|1,0,0,-,-\rangle$, $|0,0,0,+,-\rangle$ and $|0,0,0,-,+\rangle$, instead of just one in the standard STIRAP. 
However, it is still clear that if the procedure is slow enough it is possible to transfer the population adiabatically. It is also possible to switch the order of the two middle pulses \cite{multistirap}.

In this example, the population transfer takes place mainly when all four pulses differ from zero, when the product $g_{prod}=g_{1a}g_{2a}g_{1b}g_{3b}\neq0$. Letting $g_{prod}$ increase by making the pulses overlap more in time, it is possible to have efficient population transfer from state one to state five with a smaller adiabaticity parameter $G\sigma$. However, the price one has to pay is that in this case, the intermediate states become more populated during the evolution, since condition (\ref{alltimescond}) is not as well satisfied. Thus, there is a tradeoff between strict adiabaticity constraints (large $G\sigma$) and small population of intermediate states, or weaker constraints but population of the intermediate states during the transfer.

\begin{figure}
\center{\includegraphics[width=8cm,height=!]{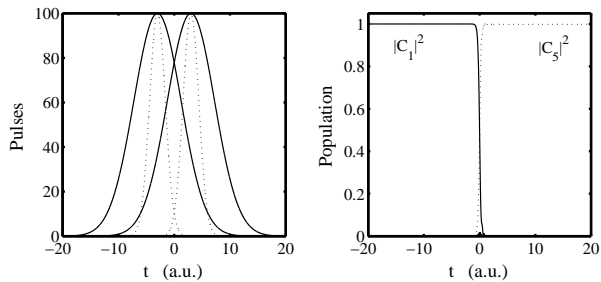}}
\caption {The same model as in figure \ref{fig2}, but with the second choice of pulse sequence, where the pulses are allowed to have different widths. The pulses, shown to the left, come in the following order: first $g_{1a}$ (solid) and $g_{3b}$ (dotted) at $t_{1a}=t_{3b}=-3$ and then $g_{1b}$ (solid) and $g_{2a}$ (dotted) at $t_{1b}=t_{2a}=3$. The widths for $g_{1a}$ and $g_{1b}$ (solid curves) are $\sigma=6$ and for the other two pulses (dotted) $\sigma=2$, and the maximum amplitudes are as in the previous example $G=100$. Time is given in units of $T$ and the pulse height in units of $T^{-1}$. The population transfer, shown in the right plot, is similar to the previous example, with a final population in cavity 3 $|C_5|^2=99.8 \%$ and a maximum population in the middle cavity equal to 0.8 $\%$.
} 
\label{fig3}
\end{figure}

Another possible coupling sequence is shown in figure \ref{fig3}. Here the coupling time between atom $a$ and cavity 1 is longer than the coupling time between and atom $b$ and cavity 3, and these two couplings are centered around the {\it same} time. Similarly, the coupling time between atom $b$ and cavity 1 is longer than the coupling time between atom $a$ and cavity 2, and these couplings are also centered around the same time. This coupling sequence also satisfies the conditions (\ref{startcond} -- \ref{alltimescond}). It could be achieved by making the diameters of laser beams 2 and 3 smaller than the diameter of laser beam 1. Numerical simulations confirm that this coupling sequence works, and the population is transferred from cavity 2 to cavity 3 with very little population of cavity 1 during the transfer. The parameters in this second choice for the pulses are, $t_{1a}=t_{3b}=-3$ and $t_{1b}=t_{2a}=3$, $\sigma_{1a}=\sigma_{1b}=6$ and $\sigma_{2a}=\sigma_{3b}=3$ and again $G=100$. The plot to the right, for the population transfer, looks similar to populations in figure \ref{fig2} and here we have final transfer in cavity 3 $|C_5|^2=99.8$ $\%$ and maximum population in the middle cavity 1 $|C_3|^2=0.8$ $\%$, thus a small fraction more than in the previous example. 

\begin{figure}
\center{\includegraphics[width=8cm,height=!]{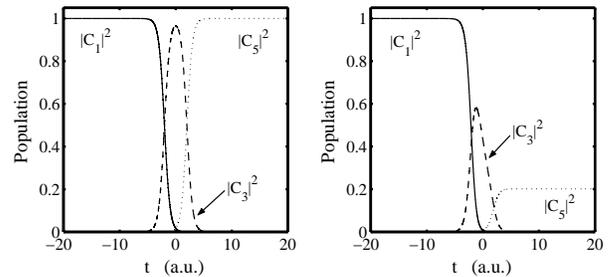}}
\caption {This figure shows the effect of losses in the intermediate cavity 1. The setup is as a double STIRAP, the first atom transfers the photon from cavity 2 to cavity 1 and finally the second atom brings it into cavity 3. Note that the pulses of the two STIRAP overlap, thus the middle cavity is never fully populated. The pulse parameters are $G=100$ (in units of $T^{-1}$),  $t_{1a}=-3$, $t_{2a}=-1$, $t_{3b}=1$ and $t_{1b}=3$ and $\sigma_{1a}=\sigma_{2a}=\sigma_{1b}=\sigma_{3b}=2$ (in units of $T$). The left plot shows the populations without losses in cavity 1, while in the right figure, cavity 1 has a decay rate $\gamma=0.1$. The final population transfer from cavity 2 to 3 is reduced from 100 $\%$ to 20 $\%$. This should be compared to, for example, using the pulse sequence of figure \ref{fig2}, with losses. If we add the same decay rate $\gamma=0.1$ for cavity 1 in that process, the population transfer goes down from 99.8 to 99.7 $\%$.   
} 
\label{fig4}
\end{figure}

Since cavity 1 remains almost  unpopulated for the coupling sequences we have discussed, relatively large losses in cavity 1 should not affect the efficiency of the state transfer. This is also confirmed by numerical simulations. 
In order to investigate the effect of losses in the intermediate cavity we add a loss term 
\begin{equation}\label{lossterm}
\delta=e^{-i\frac{\gamma t}{2}}
\end{equation}
to the derivative of the amplitude of the state $|1,0,0,-,-\rangle$.
To check the advantage of our model, without population in cavity 1, compared to a situation with population in cavity 1, we simulate a situation were atom $a$ transfers the photon first to cavity 1 from cavity 2 and then atom $b$ takes it to cavity 3. This amounts to two consecutive ordinary STIRAPs with population in the middle cavity. First we show the population transfer without losses in cavity 1 in the left plot of figure \ref{fig4} and then we add the loss term (\ref{lossterm}) to the Hamiltonian with a decay rate $\gamma=0.1$ and we see the result in the plot to the right, the transfer efficiency goes down from 100 $\%$ to 20 $\%$! When adding the same loss term to the example in figure \ref{fig2}, the decrease in population transfer is only 0.1 percentage units. The parameters for figure \ref{fig4} are $t_{1a}=-3$, $t_{2a}=-1$, $t_{3b}=1$, and $t_{1b}=3$, $\sigma=2$ and $G=100$. If we increase the decay rate to $\gamma=1$, keeping all other parameters the same, the population goes down to 99.0 $\%$ in our first method, while in the second model, when cavity 1 is populated, no population ends up in cavity 3.

Losses will, however, broaden the lineshape of the cavity. If the cavity is too long, factors $\exp(ikr)$ coming from the propagation in the cavity will most probably disturb the adiabatic transfer process, since the line is broadened and therefore not only one value, but values of $k$ in an interval are involved. For a long lossy cavity 1, the efficiency of adiabatic transfer from cavity 2 to cavity 3, trying to avoid the lossy cavity 1, will be lowered.

\subsection{Preparation of an EPR state in the "H" configuration}\label{ss33}

\begin{figure}
\center{\includegraphics[width=8cm,height=!]{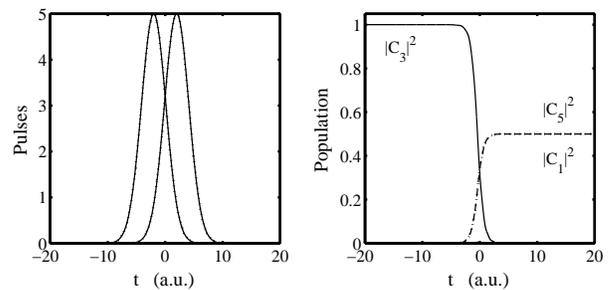}}
\caption {In this figure it is shown how well the method works for preparation of EPR states between cavity 2 and 3. To the left we show the pulses, with the parameters $G=5$ (in units of $T^{-1}$), $t_{1a}=2$, $t_{2a}=-2$, $t_{3b}=-2$ and $t_{1b}=2$and $\sigma_{1a}=\sigma_{2a}=\sigma_{1b}=\sigma_{3b}=3$ (in units of $T$). The right plot gives the populations, and it is clear that population initially in cavity 1 (solid line) is transfered equally to cavity 2 and 3 (dotted and dashed line). Note that in this situation the amplitude $G$ is much smaller than in figures \ref{fig2} and \ref{fig3}. For the fidelity in this example we have $F=|\langle EPR|\psi(t=+\infty)\rangle|=0.9999$.  
} 
\label{fig5}
\end{figure}

So far we have only been discussing transfer of a field state between two cavities, separated in space, but the model could also be used for creating entanglement between the cavities. Here we give an  example of that, and the following two sections will consider entanglement in more detail. We introduced the adiabatic eigenstate (\ref{hadiabatic}) with eigenvalue zero, and by choosing the pulses $g_{i\nu}$ carefully we could transfer population, but, of course, there are numerous other interesting pulse sequences. Assume that $g_{2a}$ and $g_{3b}$ are turned on simultaneously and then $g_{1a}$ and $g_{1b}$ are turned on simultaneously. The adiabatic state then begins, at $t=-\infty$, as $(0,0,1,0,0)^{\rm{T}}$ and ends as $(-1,0,0,0,-1)^{\rm{T}}/\sqrt{2}$. Thus, by letting atom $a$ and $b$ interact simultaneously with cavity 2 and 3 respectively, and then simultaneously interact with cavity 1, the initial photon in cavity one will be transfered into an $EPR$ state,
\begin{equation}
|EPR\rangle_{\pm}=\frac{1}{\sqrt{2}}\left(|0,1\rangle\pm|1,0\rangle\right),
\label{eprstate}
\end{equation}
of cavities 2 and 3. This procedure is shown in figure \ref{fig5}. The pulses are given in the left plot and the populations in the right plot. The parameters are $t_{2a}=t_{3b}=-2$ and $t_{1a}=t_{1b}=2$, $\sigma=3$ and $G=5$. Note that here the coupling amplitudes (and correspondingly the degree of adiabaticity) does not need to be as large as in the examples of adiabatic transfer. The photon clearly ends up in cavity 2 and 3. That the state is really the pure state (\ref{eprstate}), and not a mixture, is checked by calculating the fidelity between the final state from the numerical simulation and the $EPR$ state
\begin{equation}
F=|_+\langle EPR|\psi(t=+\infty)\rangle|.
\end{equation}
With the state obtained numerically with the parameters in figure \ref{fig5}, the fidelity becomes $F=0.9999$. By controlling the phases of the couplings it would be possible to obtain different $EPR$ states. Starting with a general field state in cavity 1, the final state would be a more complicated entangled state of cavity 2 and 3, obtained with the method explained in the previous section, by acting with the adiabatic operator $f(\hat A^{\dagger})$ on the vacuum. The situation is analogous to 
when a coherent state is split by a 50/50 beam splitter.

\subsection{"Star" configuration}\label{ss34}

\begin{figure}[!htb]
\center{\includegraphics[width=8cm,height=!]{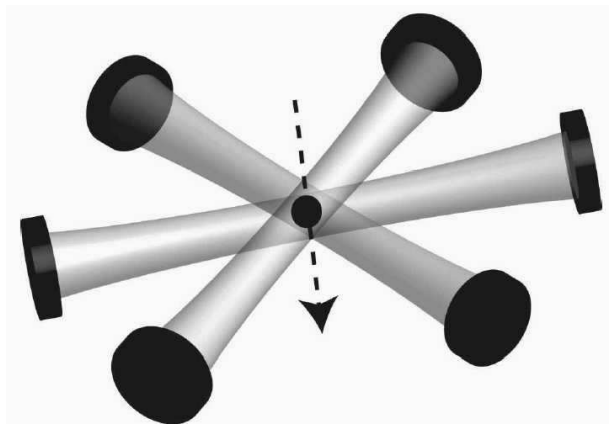}}
\caption {This figure shows a possible setup for the 'star' configuration with three cavities. Note that two of the cavities should be in the same plane, while one (the initially populated cavity) is slightly off the plane. The atom passes through the cavities in the middle point of the 'star'.
} 
\label{fig6}
\end{figure}

We can easily extend the situation to more than three cavities, or to other setups, such as a ring configuration, where the three cavities form a triangle, overlapping each others at the corners of the triangle. In this section we investigate a situation with $M$ cavities and one single atom coupled to all of the cavities, as shown in figure \ref{fig6}. We will also discuss the effect of adding further atoms coupled to some, but not all, of the cavities. If the atom travels along, say, the $z$-axis, the cavities form a "star" in the $xy$-plane. We assume that $M-1$ of them are in the same plane, centered around $z=0$,  and cavity $M$ is slightly shifted from $z=0$. Initially only cavity $M$ is populated and again we take all $\Omega_i$'s to be identical.

The effective Hamiltonian for the system is, in the rotating wave and dipole approximation, given  by
\begin{equation}\label{hstar}
H=\Delta(\sigma_a+1)+\left[g_{Ma}\hat a_M\sigma_a^++g_a\sum_{i=1}^{M-1}\,\hat a_i\sigma_a^++h.c\right].
\end{equation}
Note that we have assumed that the couplings are identical for the first $M-1$ cavities, $g_{ia}=g_a$ for $i=1,2,...,M-1$. For simplicity, we again consider only the case with one excitation, $N=1$. By labeling the states as $|1,0,..,0,-\rangle$, $|0,1,...,0,-\rangle$,...,$|0,0,...,1,-\rangle$ and $|0,0,...,0,+\rangle$, we find the adiabatic eigenstate
\begin{equation}\label{adstarstate}
|\Psi\rangle_{ad}=K( -g_{Ma}, -g_{Ma}, \ldots, -g_{Ma}, g_a, 0)^{\text T}
\end{equation}
with eigenvalue zero. Thus, if we have
\begin{equation}
\begin{array}{ccccc}
\lim_{t\rightarrow-\infty}\left(\frac{g_{Ma}}{g_a}\right)=0, & & \mbox{and} & & \lim_{t\rightarrow+\infty}\left(\frac{g_{a}}{g_{Ma}}\right)=0,
\end{array}
\end{equation}
the photon will be adiabatically transfered from cavity $M$ into all other cavities with equal probability and phase. With $M=3$, the final state in the first two cavities will be an $EPR$ state, and with $M=4$, we get a so called $W$ state,
\begin{equation}
|W\rangle=\frac{1}{\sqrt{3}}\left(|1,0,0\rangle+|0,1,0\rangle+|0,0,1\rangle\right).
\end{equation}
 For $M>4$, it is possible to prepare the natural generalization of the $W$ state to higher dimensions. A similar setup and the generation of $W$ states were discussed in \cite{w}.

\begin{figure}
\center{\includegraphics[width=8cm,height=!]{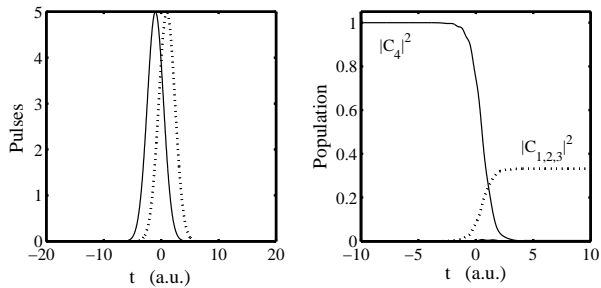}}
\caption {This shows the numerical simulation of the 'star' configuration and the preparation of a $W$ state. The left plot gives the pulses in time. The parameters are $G=5$, $\sigma_{1}=\sigma_{2}=\sigma_{3}=\sigma_{4}=2$ and $t_{1}=t_{2}=t_{3}=-1$ (dotted lines), $t_{4}=1$ (solid line). Time and the pulse widths $\sigma$ are given in units of $T$ and the pulse heights are given in units of $T^{-1}$. To the right we see the population, and it is easily seen that the initial population in cavity 4 (solid line) is equally transfered to cavities 1, 2 and 3 (dotted lines). The fidelity is $F=|\langle W|\psi(t=+\infty)\rangle|=99.8\%$.
} 
\label{fig7}
\end{figure}

In figure \ref{fig7} we show the pulses and populations during the passage of the atom, with four cavities ($M=4$). The parameters are $\Delta=0$, $G=5$, $t_{1,2,3}=1$, $t_4=-1$ and $\sigma_{1,2,3,4}=2$. The dotted lines shows the pulses $g_a(t)$ and the solid line the pulse $g_{4a}(t)$. The process is counterintuitive like the original STIRAP. In fact, this is an ``ordinary" STIRAP, but with $N-1$ final states, rather than just a single one. We clearly see that the population is equally split between the first three cavities, and with these parameters the fidelity is $F=|\langle W|\psi(t=+\infty)\rangle|=99.8$. Note that, as  for the generation of the $EPR$ state in figure \ref{fig5}, the amplitude $G$ is rather small in this example, compared to the case of population transfer between the cavities in the "H" configuration which is shown in figures \ref{fig2} and \ref{fig3}. 

\begin{figure}
\center{\includegraphics[width=8cm,height=!]{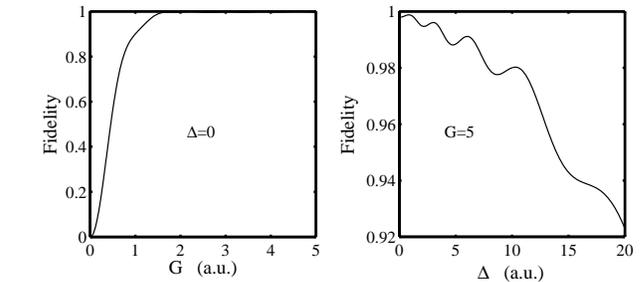}}
\caption {This figure shows the fidelity $F=|\langle W|\psi(t=+\infty)\rangle|$ as a function of the coupling-amplitude $G$ (left plot) and as a function of the detuning $\Delta$ (right plot), for the example given in figure \ref{fig7}. In the first plot $\Delta=0$ and in the second $G=5$ (in units of $T^{-1}$), otherwise the parameters are as in the previous figure \ref{fig7}.  
} 
\label{fig8}
\end{figure}

Next we show how well the process works for different parameters, changing the coupling amplitude $G$ and the detuning $\Delta$ between the atomic transition frequency $\omega_a$ and the common field frequency $\Omega$. In figure \ref{fig8}, the parameter dependence of the fidelity $F=|\langle W|\psi(t=+\infty)\rangle|$, in the previous example, is shown; first as function of the amplitude $G$, with $\Delta=0$, and then as function of $\Delta$, with $G=5$. The other parameters are as in figure \ref{fig7}. The fidelity, as expected, increases with the coupling and decreases with the detuning. Similar plots could be made for the other examples, and the information obtained would be similar.

In section \ref{s2} we explained how a general Fock state $|n\rangle$ is adiabatically transfered between two cavities. The same procedure can, of course, be used also in this configuration. In a similar fashion as in equation (\ref{adop}), we introduce an "adiabatic operator". Using the pulse sequence above, the adiabatic state (\ref{adstarstate}) will then evolve according to 
\begin{equation}
\tiny
|0,...,0,n,-\rangle\rightarrow\!\!\!\sum_{k_1+...+k_{M-1}=n}\!\!\frac{1}{N}\!\frac{n!}{k_1!...k_{M-1}!}\!\!\left(\hat a_1^{\dagger}\right)^{k_1}\!\!\!...\!\!\left(\hat a_{M-1}^{\dagger}\right)^{k_{M-1}}\!\!|0,-\rangle,
\end{equation}
where $|0,-\rangle$ on the right hand side means vacuum plus ground state atom and $1/N$ is a normalization constant. Here we have also used the multinomial theorem. Knowing how a Fock state transforms, it is easy to calculate how a general state in cavity $M$ evolves. States of similar forms as the one above, but for two modes, have been discussed for example in \cite{entangle2mod, entangle2mod2}. 
By selecting the coefficients in equation (\ref{hstar}) to differ between the individual modes, more general final states can be prepared adiabatically. 

In the adiabatic limit the system evolves according to the adiabatic states, and the process is robust against small changes in the parameters \cite{robust}, which is a great advantage for example in quantum computing \cite{adqcomp,adqcomp2}. The adiabatic states are, however, sensitive to small changes in the Hamiltonian, which will be shown next. If a second atom $b$, also in its ground state, is coupled to only cavity $j$ in the "star" configuration during the whole passage of the first atom through the $M$ cavities,  we have to add an interaction term to the Hamiltonian (\ref{hstar}) of the form
\begin{equation}\label{extra}
V=g_{jb}\hat a_j\sigma_b^++h.c.
\end{equation}  
We assume that the detuning between the $j$'th cavity and the second atom $b$ is zero, so in the interaction picture the atomic energy vanishes. The shape of $g_{jb}$ is not so important as long as it is non-zero during the process. We take it to be constant, but it could also be a very broad Gaussian, so that it extends outside the other Gaussian pulses $g_a$ and $g_{Ma}$, which could be the situation if the second atom moves much slower than the first atom and only through the $j$'th cavity. By adding the term (\ref{extra}) to the original Hamiltonian, the Hilbert space dimension obviously increases by one unit, due to the state $|0,0,..,0,-,+\rangle$, and the corresponding adiabatic state (\ref{adstarstate}) becomes
\begin{equation}
|\Psi\rangle_{ad}=K( -g_{Ma}, -g_{Ma}, \hdots, 0, \hdots, -g_{Ma}, g_a, 0, 0)^{\text T},
\end{equation}
where the new 0 is on the $j$'th position. The added atom thus  takes away the population in the $j$'th cavity. In the adiabatic limit, the magnitude of $g_{jb}$ is not important, just that it is non-zero. In other words, coupling one of the 'bare' states in the Hamiltonian weakly to a 'new' state drastically affects the adiabatic evolution. If a new atom $c$ or atom $b$ is coupled to yet another cavity $l$ during the whole interaction, the population of that state would become zero. 

\begin{figure}
\center{\includegraphics[width=8cm,height=!]{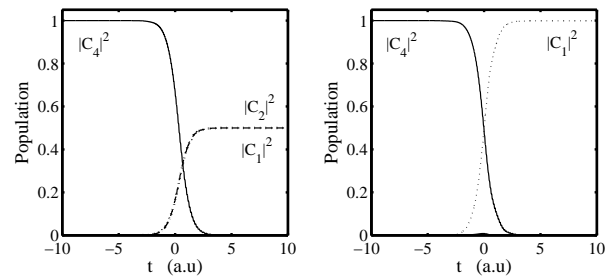}}
\caption {This figure shows the dynamics of the same 'star' configuration as in figure \ref{fig7}, but with small perturbations to the Hamiltonian. The left plot shows the same evolution as in figure \ref{fig7}, but now with a second atom $b$ coupled, in its ground state, to the cavity 3. The coupling amplitude $G$ between the first atom $a$ and the four cavities are now $G=50$, but all other parameters are the same as the previous example. The coupling between atom $b$ and cavity 3 is constant during the process, $g_{3b}=G_{3b}=5$, and the corresponding 
detuning is zero. The added atom-cavity interaction clearly modifies the evolution so that the photon ends up in cavity 1 and 2 (dashed and dotted lines). To the right we have added yet a third atom $c$, also with a constant coupling $g_{2c}=G_{2c}=5$ and zero detuning, interacting with cavity 2, and now the population in that state is removed, so that only cavity 1 is populated (dotted line). Note that atom $b$ and $c$ is much weaker coupled to the fields than atom $a$. Time is given in units of $T$ and pulse heights in units of $T^{-1}$.  } 
\label{fig9}
\end{figure}

The modification in the evolution is shown in figure \ref{fig9}. We use exactly the same example and parameters as figure \ref{fig7}, except that the common amplitude is now $G=50$. In the left plot a second atom $b$ has been coupled to the third cavity with a constant coupling $g_{3b}=G_{3b}=5$, and it is seen that all of the photon ends up in cavity 1 and 2. Note that atom $a$ is coupled ten times as strongly to the field as atom $b$. In the plot to the right, a further third atom $c$ is coupled with a constant coupling $g_{2c}=G_{2c}=5$ to cavity 2, and all population now ends up in the first state, namely the photon is in cavity 1. These plots clearly show how a small disturbance to the adiabatic Hamiltonian changes the evolution. If $G$ would have been made larger, the perturbations could have been made smaller.      

\section{State preparation using adiabatic transfer and atomic measurements}
\label{s4}

In the previous sections the atom remained more or less in its lower state during the whole process and could be seen as an ancillary state, which is never very entangled with the field state. Assuming perfect detection efficiency, a measurement on the atomic state in the $|\pm\rangle$-basis, after the interaction, would give $|-\rangle$ with unit probability. As long as the atomic state does not get entangled with the field states, an atomic measurement would not modify the cavity states. 

By introducing a third atomic level $|q\rangle$, which does not interact with the field, it is possible to create atom-field entanglement. Thus, an atom in the state $|q\rangle$ will pass through the cavities without any interaction, which could be due to a large detuning or selection rules. The Hamiltonian is correspondingly only modified by the term for the atomic energy in state $|q\rangle$, which could, of course, be omitted in a rotating frame.

In this section we will look at the "H" configuration, but other setups could also be considered. We will show how it is possible to create entangled Schr\"odinger cat states \cite{entcat, entcat1, entcat2} by measuring the atomic state after the interaction. We introduce the atomic states
\begin{equation}\label{achi}
|\chi\rangle^{a,b}_{\pm}=\frac{1}{\sqrt{2}}\left(|-\rangle^{a,b}\pm|q\rangle^{a,b}\right),
\end{equation}       
where the indices $a$ and $b$  refer to the different atoms. We will first couple cavities 1 and 2. From the STIRAP evolution
\begin{equation}
\begin{array}{c}
|0,\alpha,-\rangle\,\longrightarrow\,|-\alpha,0,-\rangle \\ \\
|0,\alpha,q\rangle\,\longrightarrow\,|0,\alpha,q\rangle
\end{array}
\end{equation}
for coherent states, it follows, starting from one of the atomic states (\ref{achi}) in the "H" configuration, that 
\begin{equation}
|0,\alpha,0\rangle|\chi\rangle_{+}^{a}\,\longrightarrow\,\frac{1}{\sqrt{2}}\left(|-\alpha,0,0\rangle|-\rangle^{a}+|0,\alpha,0\rangle|q\rangle^{a}\right).
\end{equation}
After the interaction, the atom is measured in the $|\chi\rangle_{\pm}^{a}$-basis, and depending on the measurement result the field will be in the state
\begin{equation}
N\left[|-\alpha,0,0\rangle+(-1)^{i}|0,\alpha,0\rangle\right],
\end{equation}
where $i=0$ for the measurement outcome $|\chi\rangle_+^{a}$  and $i=1$ for the result $|\chi\rangle_-^{a}$, and the normalisation constant is given by $N^{-2} =2[1+(-1)^i\exp(-|\alpha|^2)]$.

The atomic measurement in the desired basis can be effected by first using Raman pulses to couple the atomic states $|-\rangle$ and $|q\rangle$. The resulting unitary evolution should transform $|\chi\rangle_+$ into $|-\rangle$ and $|\chi\rangle_-$ into $|q\rangle$, so that  the measurement can then be implemented by testing for population in the levels $|-\rangle$ and $|q\rangle$ with a fluorescence measurement. With this procedure it is possible to reach a very high measurement efficiency, almost 100\%. Similar methods can be used to implement also generalised quantum measurements on atoms or ions \cite{atompom}.

A second atom is then injected into cavity 1 and 3 in the state $|\chi\rangle_+^{b}$. The state will evolve into
\begin{equation}
\begin{array}{c}
\frac{N}{\sqrt{2}}\left[|0,0,\alpha\rangle|-\rangle^{b}+|-\alpha,0,0\rangle|q\rangle^{b}\right.\\ \\
\left.+(-1)^{i}|0,\alpha,0\rangle|-\rangle^{b}+(-1)^{i}|0,\alpha,0\rangle|q\rangle^{b}\right].
\end{array}
\end{equation}
Atom $b$ is then measured in the same basis as that for atom $a$, with the result proportional to
\begin{equation}
|0,0,\alpha\rangle+(-1)^{j}|-\alpha,0,0\rangle+(-1)^{i}|0,\alpha,0\rangle+(-1)^{i+j}|0,\alpha,0\rangle,
\end{equation}
for the cavity field states, where $j$ is defined as $i$ is, but for atom $b$. We have here left out the normalising constant, since it will depend on the measurement outcome for atom $b$. Depending on the known measurement outcomes for atoms $a$ and $b$, we are able to prepare four possible entangled states,
\begin{equation}
\begin{array}{lll}
|\Psi_{00}\rangle\propto\left(|-\alpha,0,0\rangle+2|0,\alpha,0\rangle+|0,0,\alpha\rangle\right), & & i=j=0 \\ \\
|\Psi_{01}\rangle\propto\left(-|-\alpha,0,0\rangle+|0,0,\alpha\rangle\right), & & i=0,\,j=1 \\ \\
|\Psi_{10}\rangle\propto\left(|-\alpha,0,0\rangle-2|0,\alpha,0\rangle+|0,0,\alpha\rangle\right), & & i=1,\,j=0 \\ \\
|\Psi_{11}\rangle\propto\left(-|-\alpha,0,0\rangle+|0,0,\alpha\rangle\right), & & i=j=1.
\end{array}
\end{equation}
   
We may also consider the following scenario. If the second atom is injected in the state $|-\rangle^{b}$ instead, it will leave the setup in the same state, and the resulting field state is
\begin{equation}
N\left(|0,0,\alpha\rangle+(-1)^{i}|0,\alpha,0\rangle\right).
\end{equation}
Let us fix $\beta$ through $\alpha=2\beta$ and introduce the displacement operator $D$ with the properties
\begin{equation}
D(\beta)|\alpha\rangle=e^{i\mbox{Im}(\alpha\beta^*)}|\alpha+\beta\rangle.
\end{equation}
If the operator $D(-\beta)$ is applied to both cavity 2 and 3 and for real $\alpha$ and $\beta$, the resulting entangled state of cavities 2 and 3 becomes
\begin{equation}
N\left(|-\beta,\beta\rangle+(-1)^{i}|\beta,-\beta\rangle\right),
\end{equation}
where 
$N$ is defined as before. Here cavities 2 and 3 are both in a Schr\"odinger cat state and entangled with each other. 
This kind of entangled state is of great interest for quantum teleportation \cite{cohtel} and quantum computing with coherent states \cite{cohcomp}, but also for studying quantum phenomena in general, like entanglement and decoherence in the classical limit \cite{decoh}. Using a 50/50 beam splitter, this state may be transformed into $|\sqrt{2}\beta,0\rangle+(-1)^i|-\sqrt{2}\beta,0\rangle$, i.e. a cat state in one of the modes only, with vacuum in the other mode. 

It should be mentioned that the atomic states $|\chi\rangle_{\pm}$ could have been defined in different ways, leading to other entangled field states. The initially prepared and measured atomic basis need not be the same. We could have considered different setups of cavities and atoms and the initial coherent state could have been any state, for example squeezed states.

We conclude this section by considering another example of how to prepare a Schr\"odinger cat state. We now assume just two overlapping cavities and a single atom as in section \ref{s2}. The difference is that  the atom $a$ now should have (at least) two degenerate ground state levels $|-\rangle_{\rm I,II}$ labeled by I and II, such that the coupling amplitudes are $G_{1a,\text{I}}=G_{2a,{\rm I}}=G_{1a,{\rm II}}=G$ and $G_{2a, {\rm II}}=-G$, where 1 and 2 indicate the cavity and I,II the transition. 

One way to achieve this might be to impose a chosen quantization axis for the atom using an external electric field, thus forcing the dipole moment $\mathbf d$ of the atom to have the suitable components along the directions of the two laser fields. Alternatively it may be possible to use selection rules for the transitions in such a way, that it is possible to choose the signs of the electric field components inducing the different transitions. The choice should be made in such a way that $\mathbf d\cdot \mathbf E$ has the required signs for the four different combinations of laser and atomic transition. 

Assuming that this choice of coupling constants is possible, if we now prepare the atom in state $|-\rangle_{\rm I}^a$, an initial coherent state $|\alpha,0\rangle$ in mode 1 will be transferred into $|0,-\alpha\rangle$ in mode 2. This is because as we can see from the discussion in section \ref{ss21}, when $G_{1a,\rm I}/G_{2a, \rm I} > 0$, then an arbitrary field state $f(\hat{a}^\dagger_1)|0\rangle$ in cavity 1,  will be transferred into a state $f(-\hat{a}^\dagger_2)|0\rangle$ in cavity 2.
But if the atom is prepared in $|-\rangle_{\rm II}^a$, an initial coherent state $|\alpha,0\rangle$ in mode 1 will be transferred into $|0,\alpha\rangle$ in mode 2, {\it without} the minus sign. Again, this is because $G_{1a,\rm II}/G_{2a, \rm II} < 0$, so that an arbitrary field state in cavity 1, $f(\hat{a}^\dagger_1)|0\rangle$, will be transferred into a state $f(\hat{a}^\dagger_2)|0\rangle$ in cavity 2.
If the atom is initially in a superposition of the two states, $|\psi\rangle_\pm^a=1/\sqrt{2}(|-\rangle_{\rm I} ^a \pm|-\rangle_{\rm II}^a)$, the result will be
\begin{equation}
|\alpha,0\rangle|\psi\rangle_\pm^a\,\longrightarrow\,\frac{1}{\sqrt{2}}\left(|0,-\alpha\rangle |-\rangle_{\rm I}^a\pm|0,\alpha\rangle|-\rangle_{\rm II}^a\right).
\end{equation}
This is a Schr\"odinger cat state for cavity 2 and the atom. If we wish to disentangle the atom and the cavity, the atom may be measured in the basis $1/\sqrt{2}(|-\rangle_{\rm I}^a\pm |-\rangle_{\rm II}^a)$. Depending on the measurement outcome, we are left with one of the states
\begin{equation}
N(|0,\alpha\rangle\pm|0,-\alpha\rangle).
\end{equation}
The coherent state is transfered from cavity 1 into a cat state in cavity 2.    

\section{Conclusions}\label{s5}

In this paper, we have given several examples of cavity field state preparation and transfer using adiabatic methods. The technique we use is related to stimulated Raman adiabatic passage (STIRAP) \cite{stirap, review}. In standard STIRAP, atomic energy levels are coupled by laser pulses in order to transfer population between the atomic states. In the present scheme, cavity field mode are effectively coupled by atoms in order to transfer population between the cavity modes. A previous paper showed that not only photon number states, but arbitrary cavity field states can be transferred using this method \cite{fredrik}.
In this paper, we have in particular considered preparation of entangled states of two or more cavities, such as an $EPR$ state and a $W$ state, and various entangled superpositions of coherent states in different cavities. The theoretical considerations are supported by numerical simulations. It may also be possible to use similar techniques in solid state systems, replacing the cavities and atoms in our discussion with cavities coupled to Josephson junctions \cite{josephson}.

One advantage of adiabatic state transfer and preparation methods is that they are relatively robust against changes in the individual coupling pulse strengths and pulse durations. In contrast, state transfer e.g. in the Jaynes-Cummings model \cite{jaynescummings} relies on the ability to experimentally control the areas of coupling pulses very accurately. The situations considered in this paper are by no means totally unrealistic considering the present status of experiments in QED. An important condition is that all the cavity modes have to be degenerate. This results from energy conservation; if the modes were not degenerate, perfect state transfer between modes would not be possible. The adiabaticity for processes like the ones considered in this paper, is roughly given by the coupling amplitude times the pulse width $G\sigma$, see  \cite{fredrik}. In the example of figure \ref{fig2} we have $G\sigma=300$, while using typical experimental values of $G/2\pi\sim100$ MHz and $\sigma\sim0.3$ $\mathrm{s}^{-1}$ \cite{ent3,exppar}, the adiabaticity parameter becomes $G\sigma\approx200$. With these characteristic non-scaled parameters, the coupling is multiplied by $2\pi\cdot10^6$ and the time scales by $10^{-7}$, the adiabatic transfer of figure \ref{fig2} gives a final population of 96.9 $\%$ in the target cavity, while the maximum population in the intermediate cavity 1, during the process, is 2.2 $\%$. The whole operation, with the two atoms passing through the three cavities, takes about 2 $\mu$s, which is much shorter than the characteristic life times of cavity and the atomic states \cite{exppar}. Remember, that the example of adiabatic transfer in figures \ref{fig2} and \ref{fig3} involves more virtual intermediate levels than, for example in the generation of $EPR$  and $W$ states in figures \ref{fig5} and \ref{fig7}. Another point to emphasize is that, as the field amplitude increases, the number of intermediate states also increases, which makes the adiabaticity constrains stricter. Even though it is possible to have strong enough couplings and small decay rates for realizing the schemes proposed in this paper, it is not obvious whether it is a simple task to add crossing cavities in current experimental setups.    

In conclusion, adiabatic techniques offer rich possibilities for state transfer and population.

\section*{Acknowledgments}

We would like to thank Prof. Stig Stenholm for inspiring discussions and comments. E. A. acknowledges financial support from the Royal Society in the form of a Dorothy Hodgkin Fellowship and useful comments from Prof. Erling Riis.

\end{document}